\documentclass[aps,prb,floatfix,twocolumn,superscriptaddress]{revtex4}
\usepackage[utf8]{inputenc}
\usepackage{mathrsfs}
\usepackage{graphicx}
\usepackage{epstopdf}
\usepackage[english]{babel}
\usepackage{amsmath}
\usepackage{amssymb}
\usepackage{xcolor}
\usepackage{relsize}
\usepackage{CJK}
\usepackage{textcomp}
\usepackage{dsfont}
\usepackage{bm}
\usepackage{float}

\newcommand{\me}{\mathrm{e}}
\newcommand{\mi}{\mathrm{i}}

\newcommand{\dif}{\mathrm{d}}

\allowdisplaybreaks

\begin{document}

\title{Defect Holonomy Near Rank-Deficient Mixed States}

\author{Yu-Huan Huang}
\affiliation{School of Physics, Southeast University, Jiulonghu Campus, Nanjing 211189, China}
\author{Xu-Yang Hou}
\affiliation{School of Physics, Southeast University, Jiulonghu Campus, Nanjing 211189, China}

\author{Hao Guo}
\email{guohao.ph@seu.edu.cn}
\affiliation{School of Physics, Southeast University, Jiulonghu Campus, Nanjing 211189, China}
\affiliation{Hefei National Laboratory, University of Science and Technology of China, Hefei 230088, China}

\begin{abstract}
We investigate the geometry of mixed quantum states near rank-changing points, showing that these singularities function as effective geometric defects. The Uhlmann connection is well-defined on the full-rank sector of the density-matrix manifold, while rank-deficient states form singular boundary strata where the bundle structure degenerates. By restricting to a punctured state manifold that excludes the singular set, we obtain a well-defined gauge structure and identify an asymptotically robust invariant: the Uhlmann holonomy around noncontractible loops encircling the defect on a restricted two-dimensional punctured submanifold. In an exactly solvable qutrit model, a restricted submanifold emerges on which the connection is locally flat yet carries nontrivial monodromy, analogous to flat connections with Aharonov--Bohm-type transport. The holonomy depends only on the ratios of the vanishing eigenvalues under frozen radial dependence of the eigenbasis geometry and a fixed angular loop. In contrast, the Uhlmann curvature may diverge path-dependently when eigenvalues shrink with distinct powers, with a leading spectral-prefactor scaling law, establishing that the holonomy survives as a universal asymptotic invariant while the curvature remains non-universal. Within the effective SU(2) defect sector, the conjugacy class of the holonomy, equivalently the Wilson loop variable, provides a continuous, non-quantized classification of the asymptotic monodromy surrounding the rank-deficient defect. This non-quantization does not imply a lack of robustness: the asymptotic holonomy is an invariant of the restricted punctured submanifold and is insensitive to smooth deformations of the loop or the radial profile within the fixed spectral-ratio sector.
\end{abstract}
\maketitle

\section{Introduction}

Geometric phases are a cornerstone of topological properties in quantum systems~\cite{Bohm_GPbook,ChruscinskiBook}. The Berry phase, which arises when a quantum state undergoes adiabatic cyclic evolution in parameter space~\cite{Berry84}, has been fundamental in studying topological insulators and superconductors, where the Berry curvature serves as the central geometric object characterizing the underlying topology~\cite{TKNN,Haldane,KaneRMP,ZhangSCRMP,KaneMele,KaneMele2,ChiuRMP,Bernevigbook,BernevigPRL,MoorePRB,FuLPRL,vanderbilt2018berry,cohen2019geometric}. However, quantum systems are often in mixed states due to finite temperatures or non-equilibrium conditions, which call for generalizations of the Berry phase to mixed quantum states. A mathematically rigorous approach is the Uhlmann phase~\cite{Uhlmann86,Uhlmann89,Uhlmann1992,Uhlmann91} for mixed states. Like the Berry phase, the Uhlmann phase is obtained from parallel transport in a cyclic process, and the Uhlmann-Berry correspondence shows that the Uhlmann phase in the low-temperature limit generally agrees with the Berry phase of the ground state~\cite{P2,prq8-c9ns}. However, on any fixed-rank component the Uhlmann bundle is topologically trivial~\cite{TDMPRB15} because it admits a global section $W=\sqrt{\rho}$, so all Chern--Weil characteristic classes vanish identically. This limits its direct applicability to defining quantized topological indices within a single stratum. Nevertheless, the Uhlmann phase has been studied in various settings, including bosonic and fermionic coherent states~\cite{P2}, two-level and spin-$j$ systems~\cite{Viyuela14,P3,PhysRevA.103.042221}, and with suitable generalizations, dynamical systems~\cite{OurUhlmannQuench}. The present work shows that robust geometric information nevertheless survives at rank-changing boundaries, encoded not in local curvature integrals (Chern numbers), but in the \emph{global monodromy} of an emergent singular flat connection on the punctured manifold.

The present work is motivated by a fundamental observation: the Uhlmann bundle is defined on the space of full-rank density matrices. For a system undergoing a cyclic evolution described by $\rho(t)=\sum_i\lambda(t)|i(t)\rangle\langle i(t)|$ ($0\le t\le \tau$) with $\rho(0)=\rho(\tau)$, the Uhlmann holonomy is defined as $\mathcal{U}(C)=\mathcal{P}\exp[-\oint_C\mathcal{A}_\text{U}]$, with $\mathcal{P}$ denoting path ordering and the Uhlmann connection in its spectral form~\cite{HubnerPLA92,OurPRB20b},
\begin{align}\label{AU0}
\mathcal{A}_\text{U}=-\sum_{ij}\frac{\langle i|[\mathrm{d}\sqrt{\rho},\sqrt{\rho}]|j\rangle}{\lambda_i+\lambda_j}|i\rangle\langle j|,
\end{align}
requires strictly positive eigenvalues $\lambda_i>0$. The corresponding Uhlmann phase is $\theta_\text{U}=\arg\operatorname{Tr}[\rho(0)\,\mathcal{U}(C)]$, while the gauge-invariant Wilson loop variable is $W(C)=\operatorname{Tr}\,\mathcal{U}(C)$. At a rank-changing point $\rho_\star$ where one or more eigenvalues vanish, the spectral prefactors become ill-defined, and the Uhlmann bundle structure degenerates. The rank-deficient set therefore forms a singular boundary stratum of the density-matrix manifold, which is a stratified space where different rank sectors (with different fiber dimensions) meet.
This singular geometry is not merely a mathematical pathology but a source of profound physical effects. In a companion paper~\cite{Huang2026}, we investigated the metric side of this defect geometry: the Bures metric near rank-changing points universally develops a conical structure with a continuous deficit angle for $N\ge3$. The present work develops the complementary connection-and-holonomy side, showing that the same defect is encoded in the Uhlmann holonomy on the punctured manifold. Our guiding principle is that rank-changing points should be treated not as points where geometry simply fails, but as singularities that leave a detectable geometric imprint on the surrounding full-rank region.

The natural framework for this viewpoint is the punctured state manifold $\mathcal{M}^\times = \mathcal{M}\setminus\{\rho_\star\}$, obtained by removing the singular rank-deficient point from the state space. It is important to emphasize that the rank-deficient point itself does not belong to the full-rank Uhlmann bundle, whose base manifold consists exclusively of strictly positive density matrices. Our construction therefore does not define a holonomy at the singular point itself. Instead, the defect invariant should be understood as an asymptotic monodromy associated with loops on the punctured full-rank manifold surrounding the rank-changing point.
On $\mathcal{M}^\times$, the density matrix remains everywhere full-rank, and the Uhlmann bundle is fully well-defined. The removal of a point, when restricted to a suitably chosen submanifold, fundamentally alters the topology. The nontrivial winding structure is not a property of the full full-rank density-matrix manifold, but of the restricted two-dimensional punctured submanifold selected by the qutrit ansatz. The defect holonomy should therefore be understood as a submanifold-protected asymptotic monodromy rather than a global topological charge of the entire state space. On the restricted submanifold parametrized by $(\epsilon,\theta)$ with fixed spectral ratios and a single angular degree of freedom, loops encircling the puncture become noncontractible, with $\pi_1\simeq\mathbb{Z}$. The Uhlmann holonomy along such loops then becomes a well-posed and robust probe of the missing singularity, precisely as the Aharonov-Bohm phase encodes the magnetic flux without ever touching the solenoid. This punctured-manifold approach transforms a breakdown of the formalism into a powerful tool for diagnosing rank-deficient defects.

In this work, we construct an exactly solvable qutrit model to systematically explore this defect geometry. Our main findings are as follows. First, an emergent flat sector with nontrivial holonomy appears: on a restricted submanifold of fixed spectral ratios and a single angular degree of freedom, the Uhlmann connection is locally flat yet carries nontrivial holonomy around noncontractible loops, analogous to flat connections with nontrivial monodromy in the Aharonov-Bohm problem. Second, the defect holonomy depends only on spectral ratios: under frozen eigenbasis geometry and a fixed angular loop, the holonomy depends solely on the ratios $c_i/c_j$ of the vanishing eigenvalues, not on the radial approach to the defect or the common scaling exponent. Third, curvature singularities are path-dependent and non-universal: when eigenvalues shrink with distinct powers, the leading spectral-prefactor contribution to the Uhlmann curvature diverges with a universal power-counting form $\|\mathcal{F}_\text{U}\|\epsilon^{\Delta\alpha/2-1}$, and the singularity is path-dependent, in sharp contrast to the robust holonomy. Fourth, defect classification via Wilson loop: within the effective SU(2) defect sector, the conjugacy class of the holonomy is fully determined by the Wilson loop variable $W(C)=\operatorname{Tr}\,\mathcal{U}(C)$, yielding a continuous, non-quantized classification of rank-deficient defects. This geometric structure differs fundamentally from conventional Berry-phase topological defects. In Berry monopole problems, the topology is encoded in a curvature singularity and quantified by an integer Chern number. In contrast, the present mixed-state defect is characterized by a locally flat connection with nontrivial global monodromy on the punctured manifold. The resulting invariant is not quantized, but instead forms a continuous family of holonomy conjugacy classes determined by the asymptotic spectral composition of the vanishing subspace.

These results establish that the robust asymptotic information of a rank-deficient defect is encoded in the holonomy of an emergent flat sector on the punctured manifold, while local curvature is generally path-dependent and non-universal. The remainder of this paper is organized as follows. Section~\ref{Sec2} formalizes the geometric framework of the Uhlmann connection, the punctured manifold, and the defect monodromy assignment. Section~\ref{Sec3} presents the qutrit model, demonstrating the emergent flat connection with nontrivial holonomy, and proves the asymptotic universality of the defect holonomy, including its generalization to higher-dimensional null subspaces. Section~\ref{Sec4} introduces a two-angle extension to probe curvature singularities and establishes the path-dependent scaling theorem. We conclude and present an outlook in Section~\ref{Sec5}.

\section{Theoretical framework}\label{Sec2}
\subsection{Uhlmann connection and holonomy}

For a mixed quantum state described by a full-rank density matrix $\rho$, one can introduce the purified amplitude $W = \sqrt{\rho}\,U$, where $U \in U(N)$ is a unitary phase factor. The density matrix is recovered as $\rho = W W^\dagger$, independent of the choice of $U$. This redundancy defines a $U(N)$ principal bundle over the space of full-rank density matrices.

For a system undergoing cyclic evolution along a closed path $C$ parameterized by $\boldsymbol{R}(t)$ with $0 \le t \le \tau$ and $\boldsymbol{R}(0) = \boldsymbol{R}(\tau)$, the amplitude $W(t)$ is called a horizontal lift of $\rho(t)$ if it satisfies the Uhlmann parallel-transport condition
\begin{align}
W^\dagger \dot{W} = \dot{W}^\dagger W.
\label{parallel-transport}
\end{align}
This condition implies the differential equation $\dot{U} = -\mathcal{A}_\text{U} U$, where $\mathcal{A}_\text{U}$ is the $\mathfrak{u}(N)$-valued Uhlmann connection one-form. Upon traversing the closed loop, the initial and final purifications are related by the Uhlmann holonomy
\begin{align}
U(\tau) = \mathcal{P}\, \me^{-\oint_C \mathcal{A}_\text{U}}U(0),
\label{holonomy-def}
\end{align}
with $\mathcal{P}$ denoting path ordering. The gauge-invariant Wilson loop variable is then
\begin{align}
W(C)=\operatorname{Tr}\,\mathcal{U}(C),\qquad \mathcal{U}(C)=\mathcal{P}\exp\left[-\oint_C\mathcal{A}_\text{U}\right],
\end{align}
while the Uhlmann phase is $\theta_\text{U}=\arg\operatorname{Tr}[\rho(0)\,\mathcal{U}(C)]$.

In the spectral representation $\rho = \sum_i \lambda_i |i\rangle\langle i|$ with strictly positive eigenvalues $\lambda_i > 0$, the Uhlmann connection takes the explicit form in Eq.~(\ref{AU0}), which can be further expressed as~\cite{P2}
\begin{align}\label{AUE}
\mathcal{A}_\text{U}
=
-\sum_{i\neq j}
\frac{(\sqrt{\lambda_i}-\sqrt{\lambda_j})^2}{\lambda_i+\lambda_j}
|i\rangle\langle i|\,\dif|j\rangle\langle j|.
\end{align}
The anti-Hermiticity $\mathcal{A}_\text{U}^\dagger = -\mathcal{A}_\text{U}$ follows from $\langle i|\dif j\rangle = -\langle j|\dif i\rangle^*$ and the symmetry of the spectral prefactor under $i \leftrightarrow j$, confirming that the connection is indeed $\mathfrak{u}(N)$-valued.

\subsection{Punctured manifold and monodromy assignment}

The full space of density matrices is a stratified manifold~\cite{Bengtsson_book}, with each stratum labeled by a fixed rank.
Let $\rho_\star$ be a singular point where the rank changes.
In a sufficiently small full-rank neighborhood of $\rho_\star$, the geometry is well described by the smooth manifold $\mathcal{M}$ of full-rank density matrices.
Removing the singular point itself defines the punctured manifold
\begin{align}
\mathcal{M}^\times
=
\mathcal{M}\setminus\{\rho_\star\},
\end{align}
on which the Uhlmann bundle is everywhere well-defined.

The physical consequences of the singularity are then probed by the topology of $\mathcal{M}^\times$. The nontrivial winding structure is not a global property of the full-rank density-matrix manifold, but of the restricted two-dimensional punctured submanifold selected by the model ansatz. The defect holonomy should therefore be understood as a submanifold-protected asymptotic monodromy.
\begin{figure}[ht]
\centering
\includegraphics[width=0.9\linewidth]{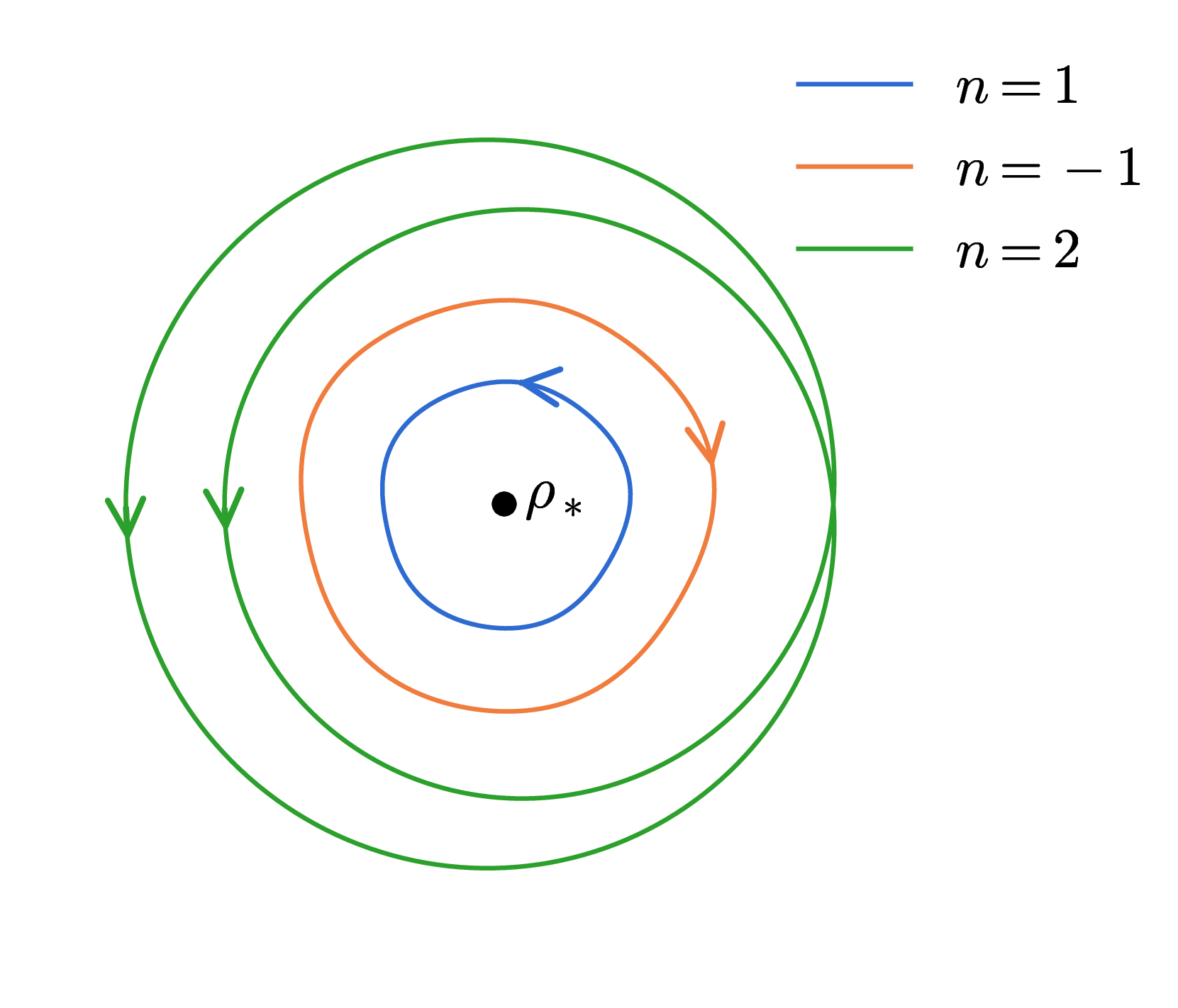}
\caption{
Schematic illustration of the punctured manifold
$\mathcal M^\times=\mathcal M\setminus\{\rho_\ast\}$.
The removed rank-deficient point $\rho_\ast$ acts as a geometric defect,
rendering loops encircling the puncture noncontractible within the restricted two-dimensional submanifold.
The three representative loops correspond to winding numbers
$n=1$, $n=2$, and $n=-1$, which generate distinct homotopy classes in
the restricted punctured disk.
}
\label{fig:puncture}
\end{figure}
On the restricted two-dimensional submanifold parametrized by $(\epsilon,\theta)$ with fixed spectral ratios and a single angular degree of freedom, this space has the topology of a punctured disk.
The removal of the point $\epsilon=0$ endows this restricted submanifold with a nontrivial topology, yielding the fundamental group
\begin{align}
\pi_1(\Sigma_\times)
\simeq
\mathbb{Z},
\end{align}
whose generator is a simple loop $C$ encircling the puncture once. As illustrated schematically in Fig.~\ref{fig:puncture}, loops in the restricted punctured submanifold are classified by the winding number $n\in\mathbb Z$.
The Uhlmann holonomy along such a loop is defined as
\begin{align}
\mathcal{U}(C) = \mathcal{P} \exp\left[-\oint_C \mathcal{A}_\text{U}\right].
\end{align}
Under a gauge transformation $U \to U h$ with $h \in \text{U}(N)$, the connection transforms as $\mathcal{A}_\text{U} \to h^{-1}\mathcal{A}_\text{U} h + h^{-1}\dif h$, from which the holonomy transforms by conjugation:
\begin{align}
\mathcal{U}(C)
\rightarrow
h^{-1}(x_0)\,\mathcal{U}(C)\,h(x_0).
\end{align}
Hence the gauge-invariant information is the conjugacy class $[\mathcal{U}(C)] \subset \text{U}(N)$. Physically, the conjugacy class characterizes the intrinsic geometric action of parallel transport around the defect, independent of the choice of local purification frame. Different holonomies related by conjugation correspond merely to different gauge choices for the local basis of purified amplitudes, while their eigenvalues, equivalently the conjugacy class, determine the observable geometric content of the defect transport. In this sense, the defect is classified by the global monodromy carried by noncontractible loops on the restricted punctured submanifold, not by a local curvature flux.
We define the defect monodromy assignment
\begin{align}
\mathfrak{D}_{\rho_\star}([C]) = [\,\mathcal{U}(C)\,],
\end{align}
which labels each homotopy class $[C]\in\pi_1(\Sigma_\times)$ by this global geometric fingerprint. In the effective SU(2) sector below, the fingerprint is fully captured by the Wilson loop
\begin{align}
W(C)=\operatorname{Tr}\mathcal{U}(C),
\end{align}
providing the direct observable signature of the defect's geometric action.

\section{Qutrit Model: Flat Connection and Defect Holonomy}\label{Sec3}

\subsection{One-parameter family and exact connection}

The general framework above is best visualized through an exactly solvable minimal example: a qutrit system with a single rank-deficient point, exhibiting a well-defined Uhlmann bundle on the punctured manifold, a locally flat yet globally nontrivial connection, and a non-quantized defect holonomy.

Consider a qutrit approaching the rank-deficient point
\begin{align}
\rho_\star = |2\rangle\langle2|.
\end{align}
We introduce the one-parameter family
\begin{align}
\rho(\epsilon,\theta)
=
\hat{U}(\theta)\Lambda \hat{U}^\dagger(\theta),
\quad
\Lambda
=
\begin{pmatrix}
\epsilon & 0 & 0 \\
0 & \zeta\epsilon & 0 \\
0 & 0 & 1-(1+\zeta)\epsilon
\end{pmatrix},
\label{rho-param}
\end{align}
where $0<\zeta<1$ and $\epsilon\to0^+$, and the unitary rotation
\begin{align}\label{U1}
\hat{U}(\theta)=\me^{-\mi\theta G},
\quad
G=\mi(|0\rangle\langle1|-|1\rangle\langle0|),
\end{align}
acts only within the nearly degenerate two-dimensional sector.
The instantaneous eigenstates are
\begin{align}
|0(\theta)\rangle&=\hat{U}(\theta)|0\rangle
=
\cos\theta |0\rangle
-
\sin\theta |1\rangle,
\notag\\
|1(\theta)\rangle&=\hat{U}(\theta)|1\rangle
=
\sin\theta |0\rangle
+
\cos\theta |1\rangle.
\end{align}
Differentiating the left-hand-sides with respect to $\theta$ gives $
|\dif0(\theta)\rangle=- |1(\theta)\rangle\dif\theta$ and $|\dif1(\theta)\rangle=
|0(\theta)\rangle\dif\theta$, leading to $\langle0(\theta)|\dif1(\theta)\rangle
=\dif\theta$ and $\langle1(\theta)|\dif0(\theta)\rangle
=
-\dif\theta$.
All other diagonal overlaps ($\langle i(\theta)|\dif i(\theta)\rangle$, $i=0,1,2$) vanish.
Substituting these overlaps into Eq.~\eqref{AUE}, only the $(0,1)$ and $(1,0)$ sectors contribute:
\begin{align}\label{AUq3}
\mathcal{A}_\text{U}
&=
-
\frac{(\sqrt{\epsilon}-\sqrt{\zeta\epsilon})^2}{\epsilon+\zeta\epsilon}
\Big(
\langle0|\dif1\rangle |0\rangle\langle1|
+
\langle1|\dif0\rangle |1\rangle\langle0|
\Big)
\notag\\
&=
-
\frac{(1-\sqrt\zeta)^2}{1+\zeta}
\Big(
|0\rangle\langle1|-|1\rangle\langle0|
\Big)\dif\theta\notag\\
&\equiv-f(\zeta)J\dif\theta,
\end{align}
where $
f(\zeta)
=
\frac{(1-\sqrt\zeta)^2}{1+\zeta}$ and $J
=
|0\rangle\langle1|-|1\rangle\langle0|$ ; we have suppressed the explicit
$\theta$-dependence of the eigenstates for convenience.
It is remarkable that the $\epsilon$-dependence cancels exactly in the spectral prefactor, leaving $f(\zeta)$ finite at the defect, a crucial property for the holonomy analysis that follows.
The operator $J$ satisfies $J^2=-I_{2\times2}$, which allows the holonomy to be summed exactly into trigonometric form below.

\subsection{Flat connection and nontrivial holonomy}

The exact solution~\eqref{AUq3} for the Uhlmann connection allows us to directly compute its curvature and holonomy, revealing the geometric structure near the defect. On the restricted submanifold with fixed spectral ratio $\zeta$, one has $\dif\zeta = 0$ and $\partial_\epsilon f =0$.
The curvature is therefore
\begin{align}
\mathcal{F}_\text{U}
=
\dif\mathcal{A}_\text{U}
+
\mathcal{A}_\text{U}\wedge\mathcal{A}_\text{U}
=0
\end{align}
since $\dif \theta\wedge \dif \theta=0$.
Thus the connection is locally flat on this restricted punctured sector. This flatness is not a trivial gauge artifact: it reflects the exact cancellation of the $\epsilon$-dependence in the spectral prefactor $f(\zeta)$, a distinctive feature of the fixed-ratio limit.

\begin{figure}[ht]
\centering
\includegraphics[width=0.49\textwidth]{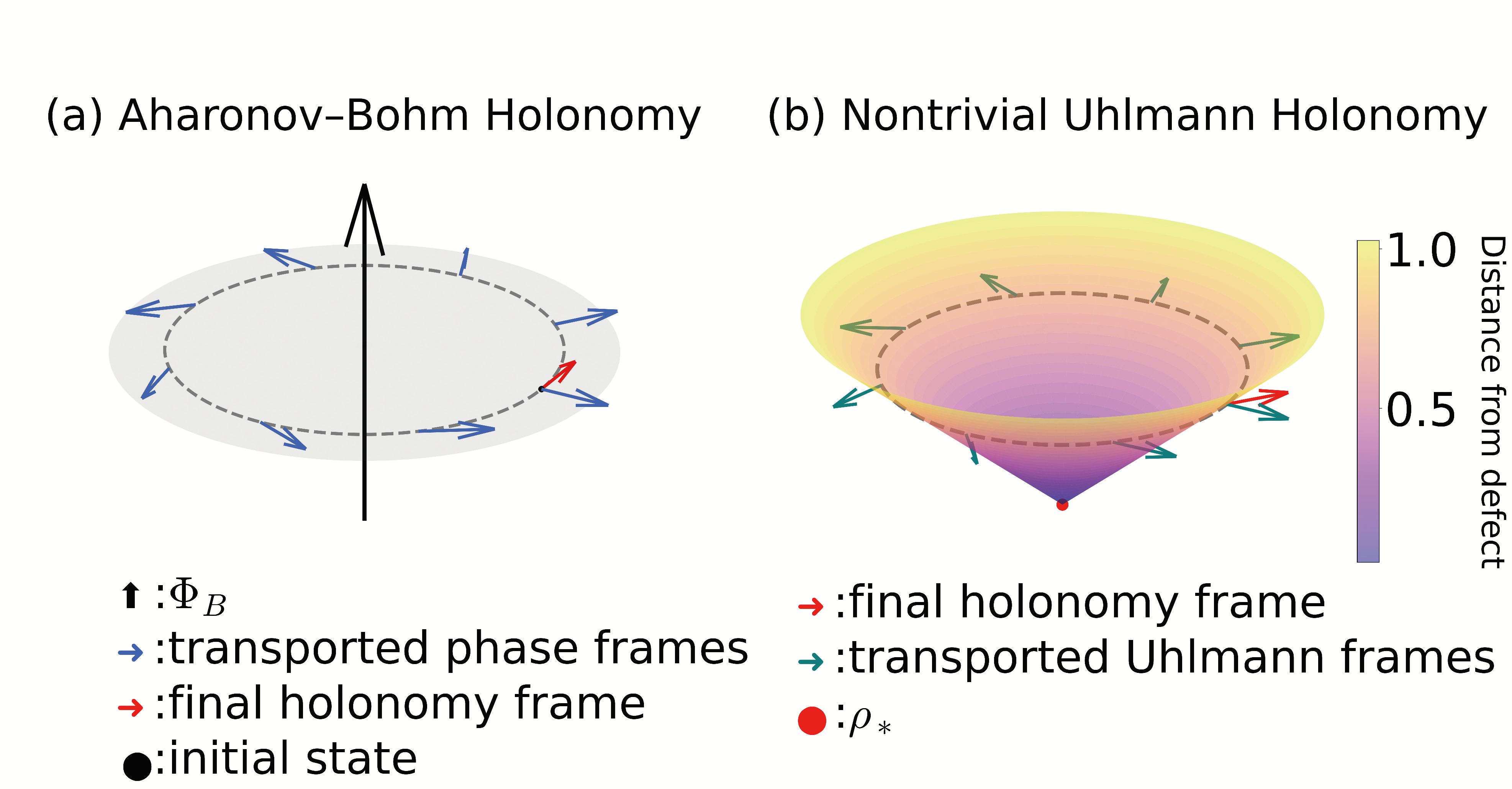}
\caption{
Comparison between the Aharonov--Bohm effect and the Uhlmann defect holonomy.
(a) Parallel transport around a magnetic flux tube in the Aharonov--Bohm effect,
illustrating the acquired geometric phase $\Phi_B$.
(b) Parallel transport on the restricted punctured submanifold
$\Sigma_\times$,
where a loop encircling the rank-deficient point $\rho_\ast$
acquires a nontrivial Uhlmann holonomy.
}
\label{fig:AB_Uhlmann}
\end{figure}

Nevertheless, the holonomy around a noncontractible loop remains nontrivial, precisely because the base manifold has a puncture. To see this, note that locally one may write
\begin{align}
\mathcal{A}_\text{U}
=
g^{-1}\dif g,
\quad
g(\theta)=\me^{-fJ\theta},
\end{align}
which can be verified by using the fact that $J$ commutes with every function of itself.
However, the gauge transformation generated by $g$ is generally multivalued around the closed loop:
\begin{align}
g(2\pi)
=
\me^{-2\pi fJ}
\neq
I
=g(0)
\end{align}
unless $f\in\mathbb{Z}$. Hence the local trivialization cannot be extended globally on the punctured manifold. This is precisely analogous to flat connections with nontrivial monodromy in the Aharonov--Bohm problem: the connection is pure gauge in any simply connected patch, yet the holonomy around the puncture is nontrivial due to the topological obstruction. The defect therefore behaves as an effective puncture-induced gauge flux: although the local curvature vanishes on the restricted manifold, the nontrivial topology of the restricted punctured submanifold obstructs the existence of a globally single-valued gauge trivialization. This geometric analogy is illustrated in Fig.~\ref{fig:AB_Uhlmann}: in both cases, parallel transport around a defect on a punctured manifold produces a nontrivial holonomy, despite the connection being locally flat away from the singularity.

The holonomy along a closed loop $C$ encircling the rank-deficient point once, parameterized by $\theta$ from $0$ to $2\pi$, is computed as
\begin{align}
\mathcal{U}(C)
=
\mathcal{P}
\exp\left[-\oint_C \mathcal{A}_\text{U}\right]
=
\exp\Big[2\pi f(\zeta)J\Big].
\end{align}
Since the connection is proportional to the constant matrix $J$, path ordering becomes trivial. Using $J^2=-I$, the exponential can be summed exactly:
\begin{align}
\me^{2\pi fJ}
&=
\sum_{n=0}^{\infty}
\frac{(2\pi fJ)^n}{n!}
=
\cos(2\pi f)I
+
\sin(2\pi f)J.
\label{U-exact}
\end{align}
Since $J$ is traceless,
we have
\begin{align}
\det\mathcal{U}(C)
=
\exp\left(2\pi f\text{Tr} J\right)
=
1.
\end{align}
Hence the holonomy belongs to SU(2). The eigenvalues are immediately found to be
\begin{align}
\me^{\pm\mi2\pi f},
\end{align}
so the gauge-invariant Wilson loop variable becomes
\begin{align}
W(C)
=
\text{Tr}\mathcal{U}(C)
=
2\cos(2\pi f(\zeta)).
\label{Wilson}
\end{align}
The defect phase
\begin{align}
\Phi_\text{U}=2\pi f(\zeta)=
2\pi
\frac{(1-\sqrt\zeta)^2}{1+\zeta}
\end{align}
varies continuously with $\zeta$, yielding a non-quantized monodromy invariant. This continuous dependence on the spectral ratio is the central result of the model: the rank-changing point acts as a geometric defect whose monodromy is not constrained by any topological quantization condition, but instead reflects the internal composition of the vanishing subspace. The dependence of the Wilson loop variable $W(C)$ on the spectral ratio $\zeta$ is shown in Fig.~\ref{fig:FUW}(a). Several representative values are collected below:
\begin{center}
\begin{tabular}{c|c|c|c}
$\zeta$ & $f(\zeta)$ & $\Phi_\text{U}$ & $W(C)$ \\
\hline
$1$ & $0$ & $0$ & $2$ \\
$7-4\sqrt3$ & $1/2$ & $\pi$ & $-2$ \\
$0$ & $1$ & $2\pi$ & $2$
\end{tabular}
\end{center}
The non-quantized nature is manifest: the phase continuously interpolates between these limits, and the Wilson loop variable takes arbitrary values in the interval $(-2,2)$. Crucially, this holonomy is an asymptotic monodromy label in the sense of the restricted punctured submanifold: the loop $C$ encircling the defect is precisely the generator of $\pi_1(\Sigma_\times)\simeq\mathbb{Z}$, and the holonomy $\mathcal{U}(C)\in\mathrm{SU}(2)$ is the image of this generator under the monodromy assignment  $\mathfrak{D}_{\rho_\star}$. Its nontrivial value reflects the fact that the flat connection on the restricted punctured disk has global monodromy obstructing its extension to the full manifold. In contrast to topological classifications fixed by integer invariants such as Chern numbers, the present defect classification is governed by a continuously tunable geometric parameter, the spectral ratio $\zeta$, because the invariant is the monodromy of a flat connection on the restricted punctured submanifold rather than a quantized curvature integral.

\subsection{Asymptotic universality and generalizations}

\subsubsection{Asymptotic spectral dependence}

The exact solution of the qutrit model reveals that the spectral prefactor $f(\zeta)$ is independent of the radial coordinate $\epsilon$. This cancellation is not a coincidence of the specific model but reflects a general asymptotic principle. To isolate this principle, we fix a family of loops $C_\epsilon:S^1\to\Sigma_\times$ surrounding the defect and assume that the overlap matrix elements
\begin{align}
\langle i(\epsilon,\theta)|\partial_\theta|j(\epsilon,\theta)\rangle
\end{align}
converge uniformly as $\epsilon\to0$.

\emph{Theorem 1 (Spectral dependence of defect holonomy under frozen geometry).}
Let $\rho(\epsilon)$ approach a rank-changing point with eigenvalues
\begin{align}
\lambda_i(\epsilon)=c_i \epsilon^\alpha+o(\epsilon^\alpha),
\qquad c_i>0,
\end{align}
and smoothly varying eigenbasis with bounded overlaps.
Fix a small loop $C$ and freeze the radial dependence of the eigenbasis geometry (i.e., the overlaps along the loop remain unchanged as $\epsilon\to0$, so that only the eigenvalues vary with the radial coordinate).
Assume the connection coefficients converge uniformly as $\epsilon\to0$.
Then the spectral dependence of the limiting holonomy enters only through the asymptotic ratios $c_i/c_j$, not through the radial profile $\epsilon$ or the common exponent $\alpha$.

\emph{Proof.}
The spectral coefficients entering Eq.~\eqref{AUE} are
\begin{align}
g_{ij}(\epsilon)
=
\frac{(\sqrt{\lambda_i}-\sqrt{\lambda_j})^2}{\lambda_i+\lambda_j}.
\end{align}
Using $\sqrt{\lambda_i(\epsilon)}=\sqrt{c_i}\,\epsilon^{\alpha/2}+o(\epsilon^{\alpha/2})$,
we obtain
\begin{align}
g_{ij}(\epsilon)
=
\frac{(\sqrt{c_i}-\sqrt{c_j})^2+o(1)}{c_i+c_j+o(1)}
\;\xrightarrow{\epsilon\to0}\;
\frac{(\sqrt{c_i}-\sqrt{c_j})^2}{c_i+c_j}.
\end{align}
Because the eigenbasis geometry is frozen along the loop, the
overlaps $\langle i|\dif|j\rangle$ are $\epsilon$-independent.
Hence the connection coefficients $\mathcal{A}_{\mathrm{U}}(\epsilon)$
converge uniformly to $\mathcal{A}_{\mathrm{U}}(0)$ on the compact loop $C$.
For bounded operator-valued functions on a compact loop, uniform convergence
guarantees termwise convergence of the Dyson series and hence commutation of
the path-ordered exponential with the limit:
\begin{align}
\lim_{\epsilon\to0}\mathcal{P}\exp\left[-\oint_C\mathcal{A}_{\mathrm{U}}(\epsilon)\right]
=
\mathcal{P}\exp\left[-\oint_C\lim_{\epsilon\to0}\mathcal{A}_{\mathrm{U}}(\epsilon)\right].
\end{align}
Therefore the asymptotic holonomy depends only on the limiting spectral
ratios $c_i/c_j$. In the qutrit model of Sec.~III.B, this is trivially
satisfied because the spectral prefactor $f(\zeta)$ becomes
$\epsilon$-independent already at finite $\epsilon$.

\subsubsection{Extension to higher-dimensional null subspaces}

The asymptotic universality established by Theorem~1 extends to
rank-changing points with a $D$-dimensional null subspace.
Consider $D$ eigenvalues vanishing with a common exponent
$\lambda_k(\epsilon)=c_k\epsilon^\alpha+o(\epsilon^\alpha)$ ($k=1,\dots,D$),
and assume the overlap matrix in the degenerate sector is driven
by a single angular coordinate $\theta$, i.e.
$\langle k(\theta)|\mathrm{d}|l(\theta)\rangle=M_{kl}\,\mathrm{d}\theta$
with constant anti-Hermitian $M$. The projected Uhlmann connection then reads
\begin{align}
\mathcal{A}_{\mathrm{U}}\big|_{\mathrm{deg}}
=-\sum_{k<l}f_{kl}\,M_{kl}\,|k\rangle\langle l|\,\mathrm{d}\theta,
\end{align}
where $f_{kl}=\frac{(\sqrt{c_k}-\sqrt{c_l})^2}{c_k+c_l}$.
The connection can be written compactly as $\mathcal{A}_{\mathrm{U}}\big|_{\mathrm{deg}}=-\tilde M(\mathbf{c})\,\dif\theta$, where $\tilde M_{kl}=f_{kl}M_{kl}$. Since $\tilde M$ is anti-Hermitian and independent of $\theta$, the holonomy around a loop $C$ is
\begin{align}
\mathcal{U}_{\mathrm{deg}}(C)
=\exp\!\bigl[-2\pi\tilde M(\mathbf{c})\bigr]\in\mathrm{SU}(D).
\end{align}
For $D=3$, a generic $\tilde M(\mathbf{c})$ has three distinct eigenvalues $\mi\phi_1,\mi\phi_2,\mi\phi_3$ with $\sum_k\phi_k=0$, yielding
\begin{align}
\mathcal{U}_{\mathrm{deg}}(C)
\sim\mathrm{diag}\!\bigl(\me^{\mi\phi_1},\me^{\mi\phi_2},\me^{\mi\phi_3}\bigr),
\end{align}
where $\phi_k=2\pi\tilde F_k(\mathbf{c})$ with $\tilde F_k$ determined by the eigenvalues of $\tilde M$. The conjugacy class is fully determined by the two complex Wilson-loop variables
\begin{align}
W_1(C)=\operatorname{Tr}\mathcal{U}_{\mathrm{deg}}(C),\quad
W_2(C)=\operatorname{Tr}\!\bigl(\mathcal{U}_{\mathrm{deg}}(C)^2\bigr),
\end{align}
which vary continuously with the spectral ratios. Thus the defect is
classified by a $(D-1)$-dimensional continuous family of holonomy
conjugacy classes, generalizing the $D=2$ result.

Eq.~\eqref{AUE} contains only off-diagonal operators $|i\rangle\langle j|$ with $i\neq j$, so its trace over the degenerate subspace vanishes identically: $\operatorname{Tr}\mathcal{A}_\text{U}\big|_{\text{deg}}=0$. Combined with anti-Hermiticity, this implies that the projected connection takes values in $\mathfrak{su}(D)$. The corresponding holonomy obeys $\det\mathcal{U}_{\text{deg}}(C)=1$, so $\mathcal{U}_{\text{deg}}(C)\in\mathrm{SU}(D)$. The conjugacy class of an $\mathrm{SU}(D)$ matrix is characterized by its $D$ eigenvalues $\{\me^{\mi\Phi_k}\}_{k=1}^D$ with $\sum_{k=1}^D\Phi_k=0\pmod{2\pi}$, or equivalently by the coefficients of its characteristic polynomial $\det(zI-\mathcal{U}_{\text{deg}})=z^D-a_1z^{D-1}+\cdots+(-1)^Da_D$, with $a_1=\operatorname{Tr}\mathcal{U}_{\text{deg}}$, $a_D=1$, and the intermediate coefficients $a_n$ expressible through the traces of powers of $\mathcal{U}_{\text{deg}}$ via the Newton identities
\begin{align}
a_n=\frac{1}{n}\sum_{m=1}^n(-1)^{m-1}a_{n-m}\operatorname{Tr}(\mathcal{U}_{\text{deg}}^m),\quad n=1,\dots,D,
\end{align}
with $a_0\equiv1$. For $D=2$, the conjugacy class is fully determined by the single real Wilson loop $W(C)=\operatorname{Tr}\,\mathcal{U}(C)=2\cos\Phi_\text{U}$. For $D\ge3$, the trace invariants $\operatorname{Tr}(\mathcal{U}_{\text{deg}}^n)$ are generally complex and subject to unitarity constraints, providing a sufficient set to determine the conjugacy class.

Importantly, the non-quantized nature persists for arbitrary $D$: the holonomy eigenvalues $\me^{\mi\Phi_k}$ vary continuously with the spectral ratios $c_k/c_l$, unconstrained by any integer invariant. The rank-changing defect is thus classified not by a discrete topological index, but by a continuous family of conjugacy classes parameterized by the asymptotic composition of the vanishing subspace.
In mathematical terms, this construction realizes a flat-connection monodromy defect: the invariant is a continuously tunable conjugacy class of the holonomy, rather than a quantized curvature integral.

\section{Path-dependent curvature singularities}\label{Sec4}

The flatness of the Uhlmann connection demonstrated in the preceding section relies on two assumptions: a fixed spectral ratio $\zeta$ and a single angular degree of freedom. Under these conditions, the connection takes the specific form $\mathcal{A}_\text{U} = -f(\zeta)J\dif\theta$, which depends on $\theta$ only through $\dif\theta$, contains no $\dif\epsilon$ component, and has an $\epsilon$-independent prefactor. These features, also encoded in Theorem~1 through the common exponent $\alpha$ for all vanishing eigenvalues, are sufficient to guarantee $\mathcal{F}_\text{U}=0$. When either assumption is relaxed, curvature singularities may emerge. To illustrate this, we simultaneously relax both assumptions by introducing a second angular coordinate $\phi$ and allowing the spectral ratio to vary with the radial coordinate, $\zeta=\zeta_0\epsilon^\delta$.
\begin{figure}[t]
\centering
\includegraphics[width=0.49\textwidth]{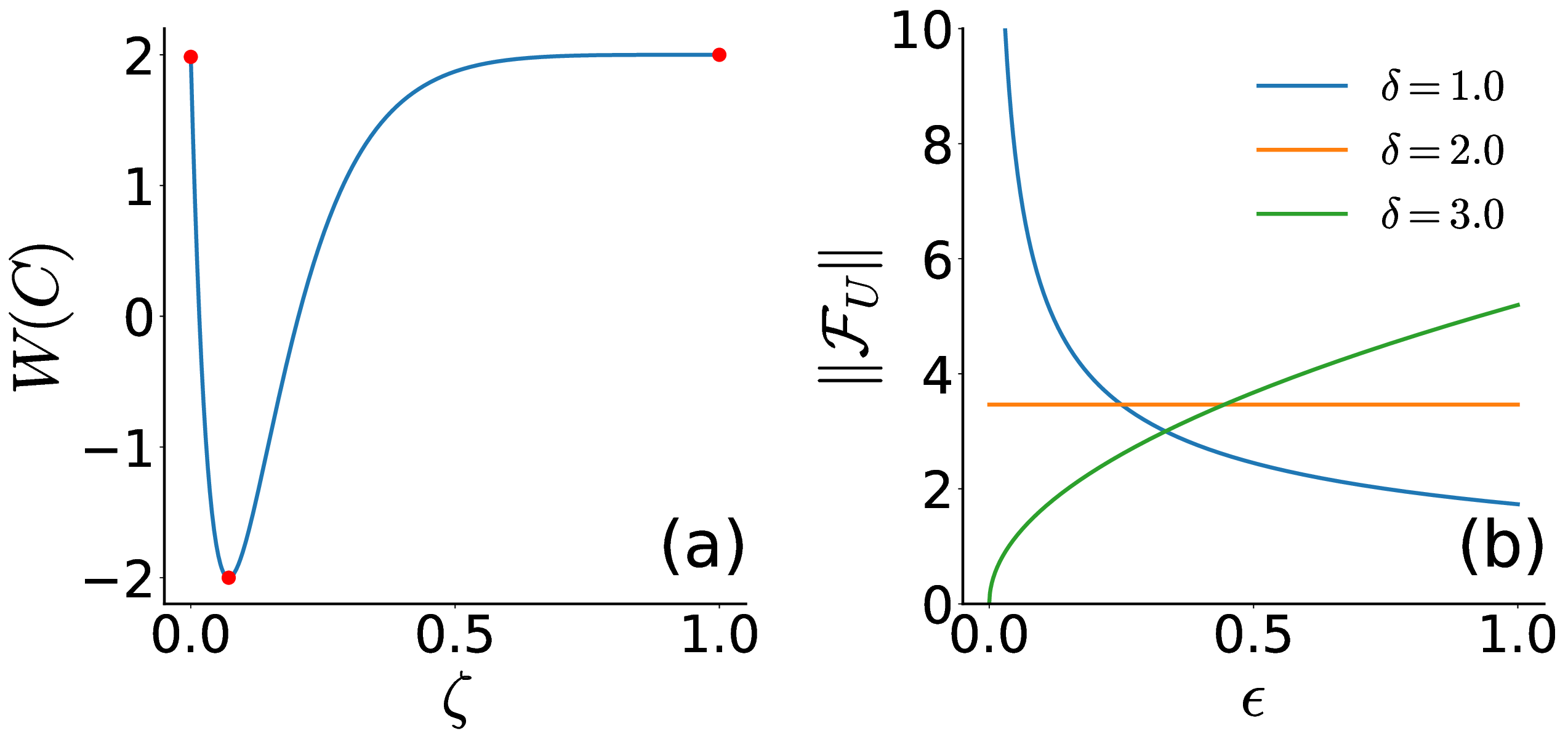}
\caption{
(a) Wilson loop variable
$W(C)=\mathrm{Tr}\,U(C)$
as a function of the spectral ratio $\zeta$.
The red points mark the representative values
$\zeta=1$, $\zeta=7-4\sqrt3$, and $\zeta=0$ discussed in the text.
(b) Leading singular scaling of the mixed curvature component
$\|\mathcal{F}_\text{U}\|_{\rm sing}\sim \epsilon^{\delta/2-1}$
for several values of $\delta$ with $\zeta_0=1$, normalized by an overall angular factor.
The three cases $\delta=1.0$, $\delta=2.0$, and $\delta=3.0$
correspond, respectively, to divergent, finite, and vanishing curvature
as $\epsilon\to0$.
}
\label{fig:FUW}
\end{figure}
The flatness result in Sec.~\ref{Sec3} was obtained with the single-parameter unitary $\hat{U}(\theta)$ in Eq.~\eqref{U1}. To probe the curvature beyond this restricted setting, we introduce a second angular coordinate $\phi$ and extend the unitary to
\begin{align}
\hat{U}(\theta,\phi)
=
\exp
\Big[
\frac{\theta}{2}
\big(
\me^{\mi\phi}|0\rangle\langle1|
-
\me^{-\mi\phi}|1\rangle\langle0|
\big)
\Big].
\label{U2}
\end{align}
When acting on the $(0,1)$ subspace, it gives
\begin{align}
|0(\theta,\phi)\rangle
&=
\cos\frac{\theta}{2}|0\rangle
-
\me^{-\mi\phi}\sin\frac{\theta}{2}|1\rangle,\notag\\
|1(\theta,\phi)\rangle
&=
\me^{\mi\phi}\sin\frac{\theta}{2}|0\rangle
+
\cos\frac{\theta}{2}|1\rangle .
\end{align}
A straightforward evaluation shows that the overlap is $\langle0(\theta,\phi)|\dif1(\theta,\phi)\rangle
=\frac{1}{2}\me^{\mi\phi}
(\dif\theta+\mi\sin\theta\dif\phi)$.
Substituting into Eq.~\eqref{AUE} and expressing the result in the Pauli basis of the degenerate subspace gives the Uhlmann connection
\begin{align}
\mathcal{A}_\text{U}
= \mathcal{A}_\theta\dif\theta+ \mathcal{A}_\phi\dif\phi.
\label{AU2}
\end{align}
with $\mathcal{A}_\theta=-\frac{\mi f(\zeta)}{2}(\sin\phi\,\sigma_x + \cos\phi\,\sigma_y)$, $\mathcal{A}_\phi=-\frac{\mi f(\zeta)}{2}\sin\theta\,(\cos\phi\,\sigma_x - \sin\phi\,\sigma_y)$. The introduction of the second angular coordinate already produces a finite angular curvature $F_{\theta\phi}$, even for fixed $\zeta$. The curvature $\mathcal{F}_\text{U} = \dif\mathcal{A}_\text{U} + \mathcal{A}_\text{U}\wedge\mathcal{A}_\text{U}$ is computed as follows. The exterior derivative gives
\begin{align}
\mathrm{d}\mathcal{A}_{\mathrm{U}} = -\frac{\mi}{2}f(\zeta)\,(\cos\theta-1)(\cos\phi\,\sigma_x - \sin\phi\,\sigma_y)\,\dif\theta\wedge\dif\phi.
\end{align}
Using $[\sigma_y,\sigma_x] = -2\mi\sigma_z$, the wedge product yields
\begin{align}
\mathcal{A}_\text{U}\wedge\mathcal{A}_\text{U}
= \frac{\mi}{2}f(\zeta)^2\sin\theta\,\sigma_z\,\dif\theta\wedge\dif\phi.
\end{align}
Summing both contributions gives the total curvature
\begin{align}
\mathcal{F}_{\mathrm{U}}
&= \frac{\mi}{2}\Big[ 2f(\zeta)\sin^2\frac{\theta}{2}\,(\cos\phi\,\sigma_x - \sin\phi\,\sigma_y) \Big]\,\dif\theta\wedge\dif\phi \notag\\
&\quad + \frac{\mi}{2}\Big[ f(\zeta)^2\sin\theta\,\sigma_z \Big]\,\dif\theta\wedge\dif\phi.
\label{FU2}
\end{align}

For fixed $\zeta$, the curvature remains finite everywhere away from the puncture. The singular behavior, however, originates from the radial derivative of the spectral prefactor when the spectral ratio becomes $\epsilon$-dependent. We now allow the spectral ratio to vary along the radial direction, taking $\zeta=\zeta_0\epsilon^\delta$ with $\delta>0$. In this case, the curvature expression~\eqref{FU2} derived from Eq.~\eqref{U2} is no longer complete: it contains only the angular component $\mathcal{F}_{\theta\phi}$ (for convenience, when referring to specific components we suppress the subscript ``U"), while the mixed radial-angular components become nonzero. According to $\mathcal{A}_\theta=-f(\zeta)\frac{\mi}{2}(\sin\phi\,\sigma_x + \cos\phi\,\sigma_y)$, $\mathcal{A}_\phi=-f(\zeta)\frac{\mi}{2}\sin\theta\,(\cos\phi\,\sigma_x - \sin\phi\,\sigma_y)$, with $\mathcal{A}_\epsilon = 0$. When $\zeta = \zeta_0\epsilon^\delta$:
\begin{align}
f(\epsilon)
=
\frac{(1-\sqrt{\zeta_0}\epsilon^{\delta/2})^2}
{1+\zeta_0\epsilon^\delta}
=
1-2\sqrt{\zeta_0}\epsilon^{\delta/2}
+O(\epsilon^\delta).
\end{align}
Differentiating yields
\begin{align}
\partial_\epsilon f
=
-\delta\sqrt{\zeta_0}\epsilon^{\delta/2-1}
+O(\epsilon^{\delta-1}).
\end{align}
For $\delta=0$, $\partial_\epsilon f=0$ identically, recovering the flat case.

Since $\mathcal{A}_\epsilon = 0$, the mixed curvature components reduce to their exterior derivative parts, and substituting $\partial_\epsilon f$ yields
\begin{align}
\mathcal{F}_{\epsilon\theta}
&=
\partial_\epsilon \mathcal{A}_\theta - \partial_\theta \mathcal{A}_\epsilon + [\mathcal{A}_\epsilon, \mathcal{A}_\theta]
\notag\\
&=
\frac{\mi}{2}\,\delta\sqrt{\zeta_0}\,\epsilon^{\delta/2-1}\,(\sin\phi\,\sigma_x + \cos\phi\,\sigma_y)+ O(\epsilon^{\delta-1}),\notag\\
\mathcal{F}_{\epsilon\phi}
&=
\partial_\epsilon \mathcal{A}_\phi - \partial_\phi \mathcal{A}_\epsilon + [\mathcal{A}_\epsilon, \mathcal{A}_\phi]
\notag\\
&=\frac{\mi}{2}\,\delta\sqrt{\zeta_0}\,\epsilon^{\delta/2-1}\,\sin\theta\,(\cos\phi\,\sigma_x - \sin\phi\,\sigma_y) \notag\\&+ O(\epsilon^{\delta-1}).
\end{align}
Both mixed components diverge as $\epsilon^{\delta/2-1}$ for $0<\delta<2$. Hence curvature singularities depend sensitively on the approach path toward the rank-changing point, in sharp contrast to the robust asymptotic holonomy established in Theorem~1. As shown in Fig.~\ref{fig:FUW}(b), the leading singular scaling of the curvature against $\epsilon$ exhibits a clear power-law dependence, with the scaling behavior governed entirely by the parameter $\delta$: for $\delta<2$ the curvature diverges near the rank-changing point, for $\delta=2$ it approaches a finite value, and for $\delta>2$ it vanishes as $\epsilon\to0$.

\subsection{Curvature scaling theorem}

The two-angle extension shows that path-dependent spectral ratios generate curvature singularities. This mechanism generalizes to any rank-changing approach where the vanishing eigenvalues do not share a common exponent, in contrast to Theorem~1 where all decaying eigenvalues scale with the same $\alpha$. We now formulate this as a scaling theorem.

\emph{Theorem 2 (Path-dependent curvature singularity).}
Let $\rho(\epsilon)$ approach a rank-changing point with eigenvalues
\begin{align}
\lambda_a(\epsilon)\sim c_a \epsilon^{\alpha_a},
\end{align}
where the exponents $\alpha_a$ are not all equal, and assume the eigenbasis varies smoothly with bounded overlaps.
Then the spectral coefficients $g_{ab}(\epsilon)$ remain bounded, while their derivatives may generate singular curvature contributions.
The leading singular contribution to the curvature originating from the spectral prefactor,
measured by the Hilbert-Schmidt norm~\cite{Bengtsson_book}
$\|\mathcal{F}_\text{U}\|:=\sqrt{\operatorname{Tr}(\mathcal{F}_\text{U}^\dagger\mathcal{F}_\text{U})}$,
scales as
\begin{align}\label{scaling-law}
\|\mathcal{F}_\text{U}\|
\sim
\epsilon^{\Delta\alpha/2-1},
\quad
\Delta\alpha
:=
\min_{a\neq b,\alpha_a\neq\alpha_b}|\alpha_a-\alpha_b|.
\end{align}
If $\Delta\alpha<2$, the contribution diverges; if $\Delta\alpha\ge2$, it remains finite.
When all relevant vanishing eigenvalues share the same leading exponent, the leading spectral-prefactor derivative vanishes. Subleading corrections may generate finite or weaker contributions, but no divergence of the form $\epsilon^{\Delta\alpha/2-1}$ arises from the leading spectral scaling.

\emph{Proof.}
Consider first a pair of eigenvalues with $\alpha_a<\alpha_b$:
\begin{align}
\lambda_a=c_a\epsilon^{\alpha_a},
\quad
\lambda_b=c_b\epsilon^{\alpha_b}.
\end{align}
The spectral coefficient becomes
\begin{align}
g_{ab}(\epsilon)
&=
\frac{\epsilon^{\alpha_a}
(\sqrt{c_a}-\sqrt{c_b}\epsilon^{(\alpha_b-\alpha_a)/2})^2}
{\epsilon^{\alpha_a}(c_a+c_b\epsilon^{\alpha_b-\alpha_a})}
\notag\\
&=
\frac{\Big(1-\sqrt{\frac{c_b}{c_a}}\epsilon^{(\alpha_b-\alpha_a)/2}\Big)^2}
{1+\frac{c_b}{c_a}\epsilon^{\alpha_b-\alpha_a}}.
\end{align}
Expanding at small $\epsilon$ yields
\begin{align}
g_{ab}(\epsilon)
&=
\Big(
1
-
2\sqrt{\frac{c_b}{c_a}}\epsilon^{\frac{\alpha_b-\alpha_a}{2}}
+
\frac{c_b}{c_a}\epsilon^{\alpha_b-\alpha_a}
\Big)
\notag\\
&\qquad\times
\Big(
1
-
\frac{c_b}{c_a}\epsilon^{\alpha_b-\alpha_a}
+
\cdots
\Big)
\notag\\
&=
1
-
2\sqrt{\frac{c_b}{c_a}}
\epsilon^{\frac{\alpha_b-\alpha_a}{2}}
+
O(\epsilon^{\alpha_b-\alpha_a}).
\end{align}
Differentiating gives
\begin{align}\label{g}
\partial_\epsilon g_{ab}(\epsilon)
\sim
(\alpha_b-\alpha_a)\epsilon^{(\alpha_b-\alpha_a)/2-1}.
\end{align}
Since these derivatives enter the curvature through $\partial_\epsilon \mathcal{A}_\text{U}$,
the leading singular scaling is controlled by the smallest nonzero exponent difference,
\begin{align}
\|\mathcal{F}_\text{U}\|
\sim
\epsilon^{\Delta\alpha/2-1},
\end{align}
which diverges for $0<\Delta\alpha<2$ and remains finite for $\Delta\alpha\ge2$.
When all relevant vanishing eigenvalues share the same leading exponent, the leading spectral-prefactor derivative vanishes. Subleading corrections may generate finite or weaker contributions, but no divergence arises from the leading spectral scaling.

To obtain the explicit Hilbert-Schmidt-norm estimate for the two-angle model, we sum over all curvature components. Using $\mathcal{F}_{\epsilon\theta}^\dagger\mathcal{F}_{\epsilon\theta}=\frac{a^2}{4}I_2$ with $a=\delta\sqrt{\zeta_0}\epsilon^{\delta/2-1}$, and similarly $\mathcal{F}_{\epsilon\phi}^\dagger\mathcal{F}_{\epsilon\phi}=\frac{a^2}{4}\sin^2\theta\,I_2$, the leading singular part of the norm squared is
\begin{align}
\|\mathcal{F}_{\mathrm{U}}\|_{\rm sing}^2
&=\operatorname{Tr}(\mathcal{F}_{\epsilon\theta}^\dagger\mathcal{F}_{\epsilon\theta})
+\operatorname{Tr}(\mathcal{F}_{\epsilon\phi}^\dagger\mathcal{F}_{\epsilon\phi})
\notag\\
&\sim \frac{a^2}{2}(1+\sin^2\theta)
= \delta^2\zeta_0\,\epsilon^{\delta-2}\frac{1+\sin^2\theta}{2}.
\end{align}
Hence $\|\mathcal{F}_{\mathrm{U}}\|_{\rm sing}\sim\delta\sqrt{\zeta_0}\,\epsilon^{\delta/2-1}$ up to an angular-dependent factor,
which is exactly Eq.~(\ref{scaling-law}) with $\Delta\alpha=\delta$. \hfill$\square$

This theorem establishes that curvature singularities are fundamentally path-dependent and therefore non-universal, in sharp contrast to the asymptotic holonomy of Theorem~1.

\section{Conclusion and Future Outlook}\label{Sec5}

We have shown that rank-changing points in mixed-state manifolds behave as effective geometric defects for which an asymptotically robust invariant emerges: the Uhlmann holonomy on the punctured state manifold. On a restricted submanifold with fixed spectral ratios and a single angular degree of freedom, the Uhlmann connection becomes flat yet exhibits nontrivial holonomy around noncontractible loops, analogous to flat connections with nontrivial monodromy in the Aharonov--Bohm problem. Theorem~1 demonstrates that, under frozen radial dependence of the eigenbasis geometry and a fixed angular loop, the spectral dependence of the holonomy enters only through the ratios of the vanishing eigenvalues, not through the radial approach or the common exponent, establishing the holonomy as an asymptotic defect invariant. When generalized to higher-dimensional null subspaces, the defect is classified by the conjugacy class of the projected holonomy, characterized by the traces of its powers via Newton identities. Theorem~2 establishes that curvature singularities originate from the spectral prefactor when eigenvalues shrink with distinct exponents, and are fundamentally path-dependent and non-universal, in sharp contrast to the robust holonomy of Theorem~1. Within the effective SU(2) defect sector, the Wilson loop variable $W(C)=\operatorname{Tr}\,\mathcal{U}(C)=2\cos\Phi_\text{U}$ provides a complete, non-quantized classification of the monodromy conjugacy class.

Together with the conical Bures metric established in prior work~\cite{Huang2026}, the present results show that these are complementary geometric signatures of the same rank-changing defect, completing a unified picture of rank-deficient mixed-state singularities.
At a rank-changing point, the geometry develops two complementary singular structures. For systems with Hilbert-space dimension $N\ge3$, the Bures metric acquires an asymptotic conical form with a continuous deficit angle, acting as a metric defect (for $N=2$ the pure-state boundary remains smooth, consistent with the Bloch-ball geometry). Concurrently, the Uhlmann connection exhibits an emergent monodromy on the restricted punctured submanifold, acting as a connection defect. Rather than being characterized by quantized local curvature invariants, these singularities are encoded through the global monodromy of a locally flat connection. The physically meaningful invariant is the conjugacy class of the asymptotic defect holonomy, a continuous geometric fingerprint that reflects the spectral composition of the vanishing subspace and is an invariant of the restricted punctured submanifold. Unlike Berry-phase topological defects whose charge is rigidly quantized by an integer Chern number, this continuous family of holonomies characterizes the defect through its parallel transport rather than through a local curvature integral. In this sense, rank-changing points behave as mixed-state analogs of monodromy defects, analogous to geometric singularities in gravity and gauge theory where curvature defects and nontrivial holonomies provide complementary characterizations of the underlying singular geometry.

An interesting future direction is the experimental detection of this defect holonomy via mixed-state interferometric protocols, such as those implemented with NMR~\cite{Zhu2011,PhysRevLett.91.100403} or superconducting qubits~\cite{Viyuela2018}, as proposed in the interferometric framework of Ref.~\cite{PhysRevLett.85.2845}. By adiabatically cycling the state around a rank-changing point in parameter space (e.g., tuning temperature or external fields), one can read off the Wilson-loop variable $W(C)$. Because Theorem~1 guarantees that the asymptotic holonomy depends only on spectral ratios and is insensitive to the radial approach, this measurement should be robust against experimental parameter drifts. Since the invariant is encoded in parallel transport around a punctured region rather than in a local curvature density, it may provide a robust probe of rank-changing structures even when local geometric quantities become strongly path-dependent.

\section{Acknowledgments}
H.G. was supported by the Quantum Science and Technology-National Science and Technology Major Project (Grant No. 2021ZD0301904) and the National Natural Science Foundation of China (Grant No. 12074064). X. Y. H. was supported by the Jiangsu Funding Program for Excellent Postdoctoral Talent (Grant No. 2023ZB611).

Yu-Huan Huang and Xu-Yang Hou contributed equally to this work.

\end{document}